\documentclass[twoside,journal]{IEEEtran}

\usepackage{graphicx}
\usepackage{fancyhdr}

\usepackage{makecell}
\usepackage{hyperref}
\usepackage{array}
\usepackage{amssymb,amsmath}
\usepackage{multicol}
\usepackage[noadjust]{cite}
\usepackage{setspace}
\usepackage{subfigure}
\usepackage{float}
\usepackage{url}
\usepackage{stfloats}
\usepackage{amsthm,pifont}
\usepackage{flushend}
\usepackage{cases,subeqnarray}
\usepackage{bm,multirow,bigstrut}
\usepackage{textcomp}
\usepackage{latexsym}
\usepackage{booktabs}
\usepackage{xcolor}
\usepackage{mathtools}
\usepackage{dsfont}
\usepackage{extarrows}
\usepackage{epsfig}
\usepackage{epstopdf}
\usepackage{algorithm}
\usepackage{algpseudocode}

\fancypagestyle{labpaperfirst}{
    \fancyhf{}
    \fancyhead[L]{%
        \includegraphics[width=4.2in]{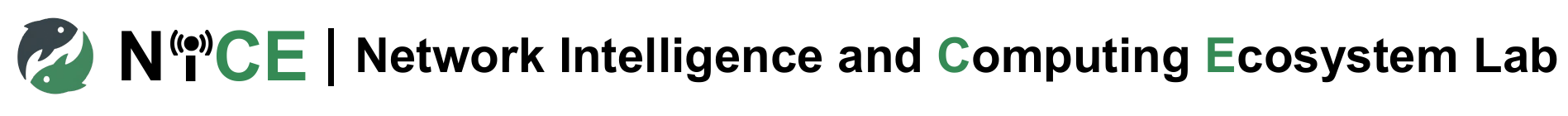}
    }

}

\theoremstyle{plain}

\begin{document}

\title{FARM: Fundamental Agentic Reward Model For
Multi-task Wireless Network Optimization}
\author{Feiran~You, Changxu Ni, Haozhe~Ma, Jun Li,~\IEEEmembership{Fellow,~IEEE}, Hongyang~Du
\thanks{F. You, H.~Du are with the Department of Electrical and Computer Engineering, The University of Hong Kong, Hong Kong SAR, China. (email: youfr@hku.hk, duhy@hku.hk) (Corresponding Author: Hongyang Du).}
\thanks{C. Ni  and J. Li  are with the National Mobile Communications Research Laboratory, Southeast University, Nanjing 210096, China (e-mail: changxuni@seu.edu.cn; jun.li@seu.edu.cn)}
\thanks{H. Ma is with Tencent Hunyuan Frontier Lab, Singapore (email: haozhe.ma@u.nus.edu)}
}
\maketitle
\thispagestyle{labpaperfirst}
\vspace{-1cm}
\begin{abstract}

Future wireless networks require learning agents to adapt across heterogeneous
channel conditions, traffic patterns, quality-of-service (QoS) requirements,
objectives, and operational constraints. Reusing decision knowledge across
such tasks is challenging because conventional multi-task and transfer
reinforcement learning methods primarily share or transfer policies, coupling
transferable knowledge with task-dependent action mappings. This paper
proposes FARM (Fundamental Agentic Reward Model for Multi-task Wireless
Network Optimization), a reward-space transfer framework that shifts
cross-task knowledge reuse from policy space to trajectory-level decision
evaluation. FARM introduces an Agentic Reward Model (ARM) that learns a
task-conditioned reward prior from heterogeneous source-task trajectories and
provides auxiliary guidance for task-specific policy optimization. In Stage~I,
ARM jointly models task conditions, temporal trajectory dependencies, and
objective-dependent reward structures while each source task retains its own
controller. In Stage~II, the learned reward prior is frozen and reused to guide
the adaptation of a target-specific controller for previously unseen tasks,
without transferring source-task policies. Experiments on heterogeneous multi-access edge computing
(MEC) tasks show that FARM achieves a mean late-stage gain of
$29.8\%$ over Single-task SAC on unseen Rate--Latency targets,
compared with $16.1\%$ for CRA Transfer, and reaches a
$46.2\%$ gain on the moderate-OOD FAR-M case.
Further analysis shows that both Mamba and Transformer trajectory encoders
support Reward-Space Transfer, while Mamba provides improved robustness
as longer history dependencies are introduced.

\end{abstract}

\IEEEpeerreviewmaketitle
\section{Introduction}

Future wireless networks are evolving from isolated optimization systems
toward intelligent and autonomous platforms, where learning-enabled agents must
continuously adapt their decisions to dynamic and heterogeneous
environments~\cite{chen2026toward}. Unlike conventional wireless optimization,
which typically assumes a predefined objective under a fixed system
configuration, future networks must accommodate diverse users, services, and
operational requirements under varying channel conditions, traffic patterns,
quality-of-service (QoS) preferences, and optimization
objectives~\cite{jiao2025intelligent}. Multi-access edge computing (MEC)
provides a representative setting for such intelligent wireless control, where
communication, computation, and energy resources must be jointly optimized
under complex and time-varying conditions~\cite{ullah2025nodes}. Accordingly, future wireless intelligence requires not only effective
decision-making for individual tasks, but also the ability to acquire and reuse
optimization knowledge across heterogeneous tasks.
Instead of independently learning a controller for every task, scalable
wireless intelligence requires mechanisms that can reuse and adapt optimization
knowledge across related yet distinct task distributions~\cite{ghafouri2026toward}.
These tasks may differ in channel statistics, traffic patterns, QoS priorities,
constraint configurations, and optimization preferences, such as throughput
maximization, latency reduction, energy minimization, or their combinations.
Effective cross-task knowledge reuse therefore becomes a central challenge for
intelligent wireless networks.

Deep reinforcement learning (DRL) has demonstrated strong capability for
adaptive wireless optimization through interaction with dynamic
environments~\cite{11623419}. However, most existing DRL approaches remain
task specific, with learned policies closely coupled to particular
environments, objectives, and action mappings. Multi-task reinforcement
learning (MT-RL) and transfer learning improve knowledge reuse primarily
through policy sharing or policy transfer. Although effective for related
tasks, such policy representations inherently entangle reusable optimization
knowledge with task-dependent action generation. When channel statistics,
objectives, QoS requirements, or constraints change, transferred policies may
retain source-specific control patterns and require substantial adaptation.
This coupling can introduce cross-task interference and limit the scalability
of policy-level transfer under heterogeneous wireless conditions. This limitation motivates a fundamental question: \emph{what should be
transferred across heterogeneous wireless optimization tasks?} Therefore, we investigate reward
representations as an alternative interface for cross-task knowledge reuse.
Our key premise is that, although optimal actions may differ substantially
across tasks, the principles used to evaluate decision trajectories can exhibit
stronger cross-task regularities. Such trajectory-level evaluation knowledge
captures how sequences of observations and actions contribute to long-term
optimization objectives under different wireless conditions without directly
specifying a task-dependent action mapping. This observation motivates
\emph{Reward-Space Transfer}, in which transferable reward representations are
learned from heterogeneous interaction trajectories and reused as auxiliary
guidance for task-specific policy optimization.

Realizing effective Reward-Space Transfer raises three key challenges.
First, heterogeneous wireless tasks have different evaluation semantics due to
variations in optimization objectives, QoS requirements, channel conditions,
and constraints. The same trajectory may therefore have different utility
under different task contexts, requiring reward evaluation to be explicitly
conditioned on task characteristics. Second, reward learning must balance
cross-task commonality and task-specific specialization: a fully shared
representation may suppress objective-dependent preferences, whereas
independent reward models eliminate the benefit of knowledge reuse. Third,
reward knowledge learned from source tasks must remain useful under unseen
task configurations. Since target tasks may introduce new combinations of
channel dynamics, QoS requirements, objectives, and constraints, transfer must
selectively exploit relevant prior knowledge while avoiding mismatched reward
guidance under distribution shift.

\begin{figure*}[t]
\centering
\includegraphics[width=0.75\textwidth]{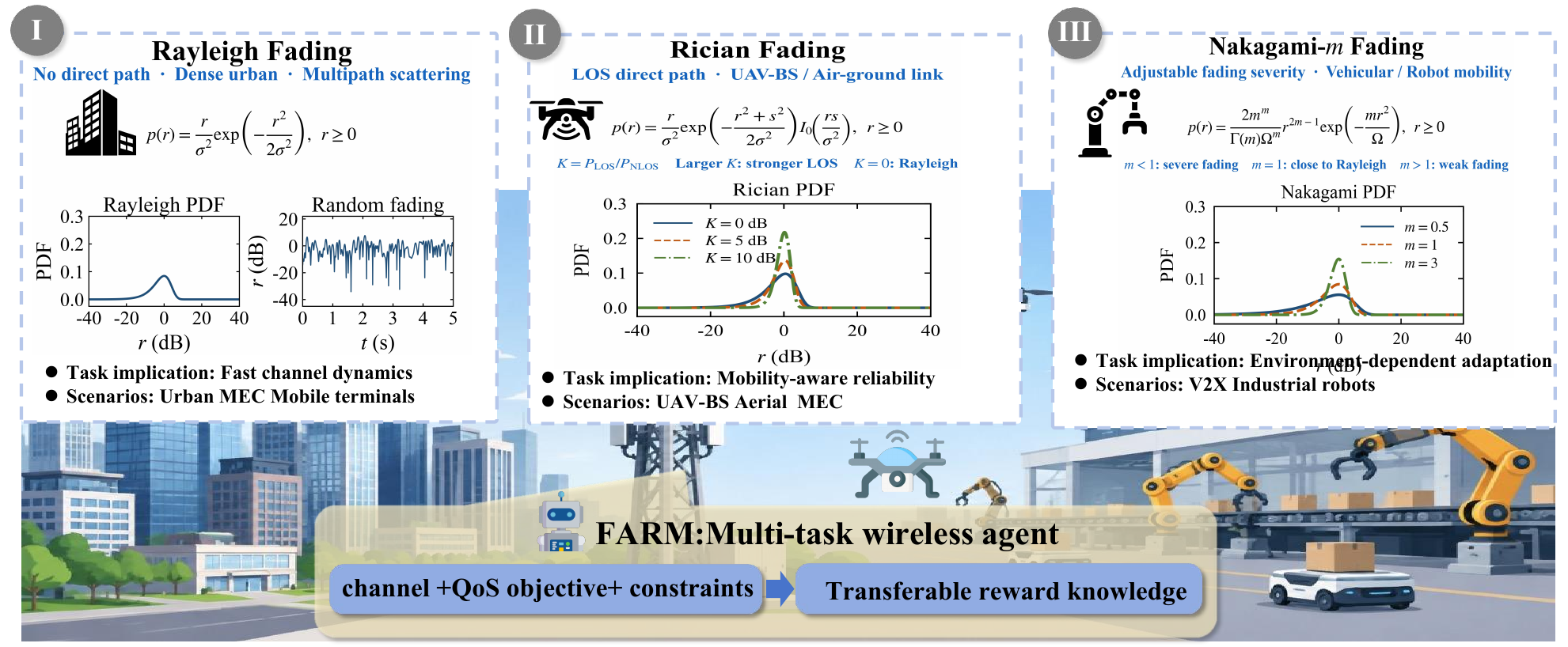}
\caption{
Overview of heterogeneous wireless tasks considered by FARM.
The task space spans diverse channel conditions, QoS objectives, constraints,
and deployment scenarios. (I) Rayleigh fading models dense urban and mobile
MEC environments with rapid non-line-of-sight variations. (II) Rician fading
represents line-of-sight-dominant UAV-assisted and air-ground MEC scenarios.
(III) Nakagami-$m$ fading captures adjustable propagation severity in V2X and
industrial robotic environments. These heterogeneous tasks support
trajectory-level reward learning for Reward-Space Transfer.
}
\label{fig_main}
\end{figure*}

To address these challenges, we propose FARM, a Fundamental Agentic Reward
Model for multi-task wireless network optimization, as illustrated in
Fig.~\ref{fig_main}. FARM shifts the
transferable component from task-specific policies to trajectory-level
decision evaluation knowledge. Its core component, the Agentic Reward Model
(ARM), learns a transferable reward prior from heterogeneous wireless
interaction trajectories. Importantly, the learned reward does not replace the
original task objective; instead, it provides auxiliary guidance that
complements the environment reward during task-specific policy optimization.
FARM therefore separates reusable reward knowledge from task-dependent action
generation while allowing individual controllers to remain specialized to
their corresponding wireless environments. FARM follows a two-stage learning and transfer paradigm. In Stage~I, ARM learns
a transferable reward prior from a heterogeneous source-task bank by jointly
modeling task conditions, temporal trajectory dependencies, and
objective-dependent reward structures, while each source task retains its own
controller. In Stage~II, the learned reward prior is frozen and reused for
previously unseen target tasks without transferring source-task policies.
Objective-aware routing and task-conditioned reward generation provide
auxiliary guidance, while the target-specific controller is optimized using
the original environment reward together with the transferred reward signal.

To support task-aware reward reasoning, ARM combines a task-condition interface
for channel configurations, optimization preferences, QoS requirements, and
constraints with a trajectory representation module over historical
observations, actions, and rewards. Condition-aware feature modulation and
objective-specific reward experts reconcile shared reward knowledge with
task-dependent evaluation semantics, while adaptive auxiliary-reward weighting
controls the contribution of transferred guidance. A Mamba-based selective
state-space encoder is adopted as the default temporal backbone to capture
long-range dependencies in dynamic wireless environments. Together, these
components enable ARM to learn reusable trajectory-level reward knowledge
while preserving the task specificity required for heterogeneous wireless
optimization.

The main contributions of this paper are as follows:

\begin{itemize}

\item We introduce FARM, a Fundamental Agentic Reward Modeling framework that
formulates \emph{Reward-Space Transfer} for heterogeneous wireless
optimization. Unlike policy sharing or policy transfer, FARM transfers
trajectory-level decision evaluation knowledge and uses the learned reward
prior as auxiliary guidance for task-specific policy optimization, thereby
decoupling transferable knowledge from task-dependent action mappings.

\item We develop ARM for learning transferable reward representations from
heterogeneous wireless trajectories. ARM integrates task-condition encoding,
temporal trajectory modeling, condition-aware feature modulation,
objective-specific reward experts, and adaptive auxiliary-reward weighting to
capture reusable evaluation knowledge across tasks while retaining task-specific
objective and constraint information.

\item We design a two-stage reward-space learning and transfer mechanism in
which a transferable reward prior is learned from heterogeneous source tasks
and subsequently reused on unseen target tasks while the shared reward prior
remains frozen. Objective-aware routing and task-conditioned reward guidance
support target-policy adaptation under distribution shifts. Experiments on
heterogeneous MEC tasks demonstrate effective multi-task reward learning and
Reward-Space Transfer to unseen tasks, while temporal-backbone analysis shows
that Mamba provides improved robustness when longer history dependencies are
introduced.

\end{itemize}


The remainder of this paper is organized as follows: Section~\ref{related} discusses related work. Section~\ref{sys} presents the system model. Section~\ref{problem} gives the problem formulation. Section~\ref{algorithm} details the design of the algorithm. Section~\ref{num} presents experimental results and analysis. Finally, Section~\ref{conclusion} concludes the paper with a summary of our key findings.

\section{Related Work}
\label{related}

\subsection{Multi-Objective Optimization in Wireless Networks}

Wireless resource management commonly involves conflicting objectives such as
throughput, latency, and energy efficiency. Classical methods typically rely on
scalarization, including weighted-sum and $\epsilon$-constraint
formulations~\cite{eichfelder2008adaptive}, and have been applied to power
control~\cite{ding2025interference}, subcarrier
allocation~\cite{thet2024user}, and user scheduling~\cite{rai2024spectral}.
Pareto-oriented methods based on evolutionary optimization or successive convex
approximation have also been investigated~\cite{tartibu2025multi}. These
approaches are effective under well-specified models but generally rely on
explicit objective formulations and system assumptions, which limits their
flexibility under changing wireless conditions and task requirements.

\subsection{DRL for Wireless Resource Management}
DRL provides a model-free alternative for adaptive wireless resource
management. Single-agent methods have been applied to power control, rate
adaptation, and caching~\cite{rios2025leveraging}, while multi-agent methods
address distributed interference management and device-to-device
communication~\cite{wang2024scheduling}. SAC and its variants are 
attractive because of entropy-regularized exploration~\cite{xu2025enhancing},
and wireless constraints can be incorporated through constrained Markov
decision processes and Lagrangian formulations
\cite{sun2025comprehensive}. Most existing approaches,
however, optimize task-specific rewards and policies, making the learned
decision rules closely dependent on particular objectives, QoS requirements,
and network conditions.

\subsection{Multi-Task RL for Wireless Networks}

MT-RL improves generalization by sharing policies
or value representations across related tasks~\cite{chen2025multi,zhu2024multi}.
Parameter sharing and context conditioning have been used across environments
and user configurations, including cross-device and
multi-scenario wireless optimization. Meta-RL
methods  target rapid task adaptation through MAML and related
approaches~\cite{zhao2025meta}. Sequence modeling provides another
direction, where trajectory histories are used for multi-task decision making.
WiSDoM, for example, combines a sparse Decision Transformer with
Mixture-of-Experts for heterogeneous mobile-network
optimization~\cite{temiz2026wisdomwirelesssparsedecision}. These methods
primarily transfer policy, value, or decision representations, so transferable
knowledge remains closely coupled with action generation.

\subsection{Reward Shaping and Auxiliary Reward Modeling}

Reward shaping incorporates prior knowledge into the learning
signal~\cite{you2025dress}, while auxiliary reward models can
approximate complex utilities or provide denser feedback in multi-objective
RL~\cite{ma2024reward}. Related learned surrogate models have also been used in
wireless applications such as channel prediction and beamforming
optimization~\cite{li2025joint}. CRA is particularly relevant to our setting because it introduces a
centralized reward agent that provides auxiliary guidance to task-specific
policies~\cite{NEURIPS2025_eeb273b2}. Its teacher generates auxiliary rewards
from instantaneous state--action pairs and combines them with the environment
reward through a shared scaling mechanism. This improves reward densification
but does not explicitly model task-conditioned trajectory semantics or reward
transfer across tasks with simultaneously varying objectives, QoS
requirements, channel dynamics, and constraints.

\begin{table*}[!t]
\centering
\footnotesize
\setlength{\tabcolsep}{3pt}
\renewcommand{\arraystretch}{0.98}
\caption{Main notations.}
\label{tab:notations}

\begin{tabular}{@{}lp{4.3cm}lp{4.15cm}lp{4.5cm}@{}}
\toprule
Symbol & Definition
& Symbol & Definition
& Symbol & Definition \\
\midrule

\multicolumn{6}{@{}l}{\textit{Task and wireless-system notation}} \\

$N$
& Number of users
& $M$
& Number of source tasks
& $H$
& Episode horizon \\

$k$
& User index
& $\tau$, $\tau_i$
& Wireless task and source task
& $\mathcal{B}$
& Source task bank \\

$\mathcal{H}_{\tau}$
& Channel configuration
& $\lambda_{\tau}$
& Traffic intensity
& $\bm{q}_{\tau}$
& QoS thresholds \\

$\mathcal{C}_{\tau}$
& Active constraint set
& $\bm{w}_{\tau}$
& Objective weights
& $c$, $d_c$
& Task-condition vector and dimension \\

$B$, $N_0$
& Bandwidth and noise power
& $\rho_{\tau}$
& Temporal correlation coefficient
& $h_{k,t}$
& Channel power gain \\

$p_{k,t}$
& Transmit power
& $\beta_{k,t}$
& Offloading ratio
& $P_{\mathrm{tot}}$
& Total transmit-power budget \\

\midrule
\multicolumn{6}{@{}l}{\textit{MEC computation and performance notation}} \\

$L_{k,t}$
& Input data size
& $\xi_{k,t}$
& CPU cycles per bit
& $P_k^{\mathrm{loc}}$
& Local computation power \\

$f_k^{\mathrm{loc}}$
& Local CPU frequency
& $f^{\mathrm{edge}}$
& Edge CPU frequency
& $R_{k,t}$
& Achievable rate \\

$D_{k,t}$
& Completion latency
& $E_{k,t}$
& Energy consumption
& $\eta_{k,t}$
& Energy efficiency \\

$U_{\tau}$
& Task-specific objective
& $r_{\mathrm{env}}$
& Environment reward
& $v_t$
& Aggregated normalized violation cost \\

$\mathcal{S}$, $\mathcal{A}$
& State and action spaces
& $P_{\tau}$
& Transition probability
& $\gamma$
& Discount factor \\

$o_t$, $a_t$
& Observation and action
& $\lambda_{\mathrm{pen}}$
& Constraint penalty coefficient
& $U_{\tau}^{\mathrm{ref}}$
& Task-specific normalization reference \\

\midrule
\multicolumn{6}{@{}l}{\textit{Reward-prior and transfer notation}} \\

$z_t$
& Trajectory token
& $\Gamma_t$
& Trajectory window
& $W$
& History-window length \\

$\psi_t$
& Reward representation
& $\hat r_t$
& Auxiliary reward
& $\lambda_{\mathrm{aux}}(c)$
& Auxiliary-reward weight \\

$\mathrm{Expert}_g$
& Objective-specific reward expert
& $\mathrm{CondLam}(\cdot)$
& Conditional reward weighting
& $\Theta$
& Reward-prior parameters \\

 $\mathcal{L}_{\mathrm{reward}}$
& Reward-prior learning objective
& $\pi_i$, $\pi_{\mathrm{tar}}$
& Source- and target-task policies\\

\bottomrule
\end{tabular}
\end{table*}
\section{System Model}
\label{sys}

\begin{figure}[t]
\centering
\includegraphics[width=0.5\textwidth]{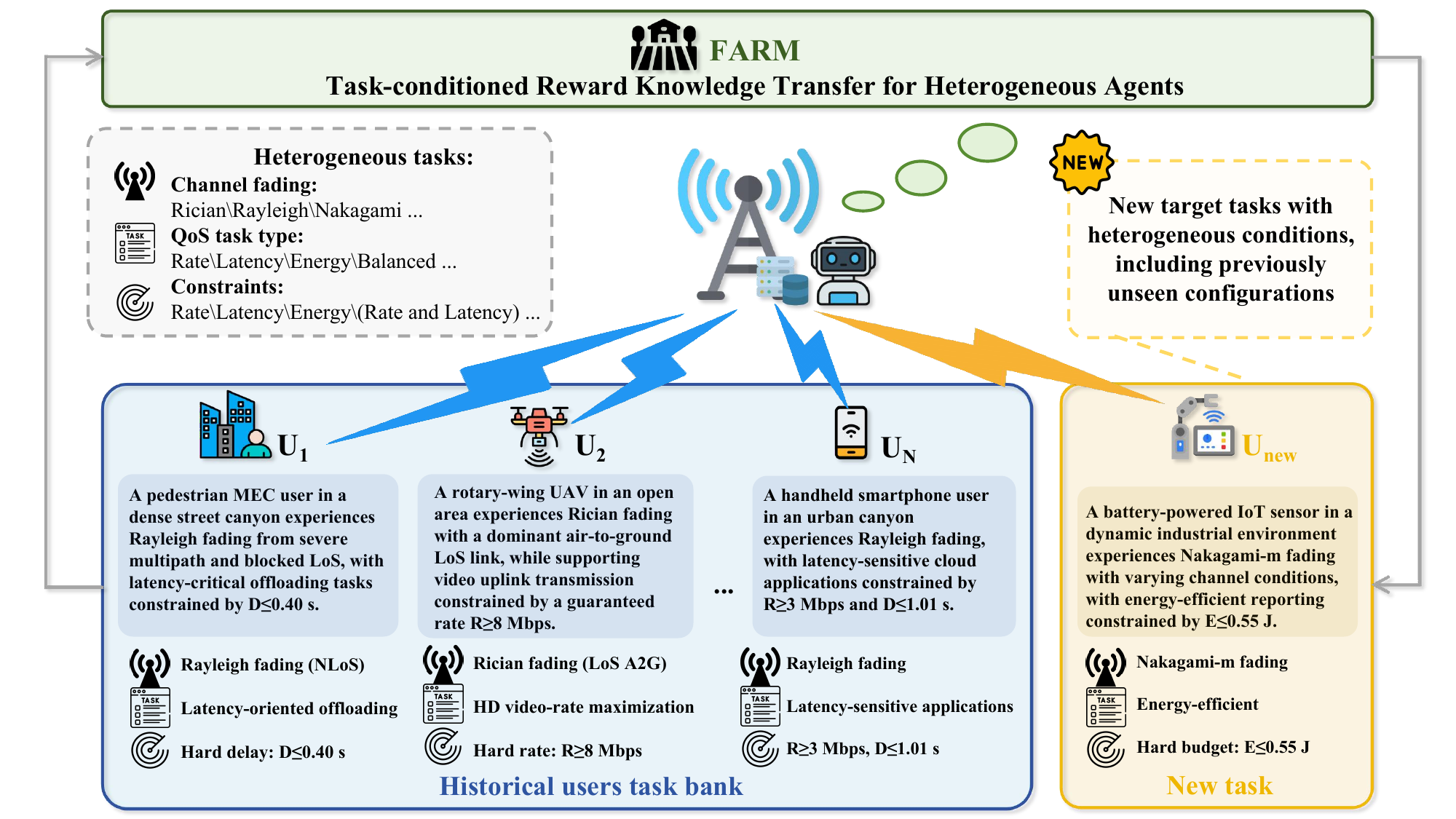}
\caption{
Heterogeneous multi-task MEC system with Reward-Space Transfer.
Task heterogeneity arises from variations in channel dynamics, traffic,
QoS requirements, objectives, and operational constraints. For unseen tasks
or task configurations, FARM reuses the learned reward prior to provide
task-conditioned guidance for target-specific policy adaptation.
}
\label{system}
\end{figure}
We consider a representative uplink MEC domain in which $N$ wireless users are served by one base station (BS) with an edge server. The per-BS model is used as the basic decision domain and does not restrict FARM to single-cell deployment. 
In a multi-BS network, FARM can be instantiated at each BS for the heterogeneous tasks within its coverage, since the transferable reward prior is task-conditioned. 
The system operates over discrete time slots $t\in\{0,1,\ldots,H-1\}$. At each slot, the controller observes the wireless and computation states and decides transmission power and computation offloading schemes for the served users. Each task corresponds to a specific MEC configuration characterized by its channel conditions, traffic conditions, QoS requirements, resource constraints, and optimization objective. Variations in these factors define the heterogeneous task space for Reward-Space Transfer.

\subsection{Network and Channel Model}
\label{sec:channel}

Let $h_{k,t}>0$ denote the instantaneous channel power gain of user $k$ at
time slot $t$. To characterize heterogeneous wireless environments, different
small-scale fading conditions are considered, including Rayleigh fading,
Rician fading, and Nakagami-$m$ fading. Rayleigh fading represents rich
non-line-of-sight multipath propagation, where the channel is dominated by
scattered components~\cite{marins2019fading}. Rician fading captures the
existence of a dominant line-of-sight component through the Rician factor
$K_{\tau}$~\cite{le2026performance}, while Nakagami-$m$ fading provides a
flexible representation of different propagation conditions through the shape
parameter $m_{\tau}$ and average power $\Omega_{\tau}$
~\cite{badarneh2024fluctuating}. Accordingly, the channel configuration associated with task $\tau$ is defined
as
$
\mathcal{H}_{\tau}
=
(\mathrm{Family}_{\tau},
\sigma_{\tau},
K_{\tau},
m_{\tau},
\Omega_{\tau}),
$
where $\mathrm{Family}_{\tau}$ denotes the fading family and the remaining
parameters characterize the corresponding fading severity. Different channel
configurations induce different wireless task distributions and provide the
basis for evaluating transferable reward representations under heterogeneous
environments. To capture temporal channel variations, the channel evolution is modeled by a
first-order autoregressive process
$
h_{k,t+1}
=
\rho_{\tau}h_{k,t}
+
(1-\rho_{\tau})\tilde h_{k,t+1},
$
where $\tilde h_{k,t+1}$ is independently sampled from the task-specific
fading distribution and $\rho_{\tau}\in[0,1)$ controls the temporal
correlation. A smaller $\rho_{\tau}$ indicates faster channel variation and a
more dynamic wireless environment.

Given the transmit power $p_{k,t}$ of user $k$, the achievable data rate is
calculated as
\begin{equation}
\label{eq:shannon}
R_{k,t}
=
\frac{B}{10^{6}}
\log_2
\left(
1+
\frac{p_{k,t}h_{k,t}}{N_0}
\right),
\end{equation}
where $R_{k,t}$ is measured in Mbps.

\subsection{Computation Offloading and Resource Consumption Model}
\label{sec:offload}

Each user generates computation-intensive tasks that can be processed locally
or offloaded to the edge server. Let $L_{k,t}$ denote the input data size of
user $k$ at time slot $t$, and let $\xi_{k,t}$ represent the required CPU
cycles per bit. The computation offloading decision is characterized by
$\beta_{k,t}\in[0,1]$, where $\beta_{k,t}$ denotes the fraction of input data
offloaded to the edge server. Let $f_k^{\mathrm{loc}}$ and $f^{\mathrm{edge}}$ denote the local CPU
frequency of user $k$ and the computing frequency of the edge server,
respectively. The local execution latency is given by
\begin{equation}
\label{eq:local_delay}
D_{k,t}^{\mathrm{loc}}
=
\frac{
L_{k,t}(1-\beta_{k,t})\xi_{k,t}
}
{
f_k^{\mathrm{loc}}
}.
\end{equation}
For offloaded tasks, the completion latency consists of transmission latency
and edge execution latency:
\begin{equation}
\label{eq:edge_delay}
D_{k,t}^{\mathrm{edge}}
=
\frac{
L_{k,t}\beta_{k,t}
}
{
R_{k,t}10^{6}
}
+
\frac{
L_{k,t}\beta_{k,t}\xi_{k,t}
}
{
f^{\mathrm{edge}}
}.
\end{equation}
Since local and edge execution can proceed in parallel, the overall completion
latency is determined by
\begin{equation}
\label{eq:latency}
D_{k,t}
=
\max
\left\{
D_{k,t}^{\mathrm{loc}},
D_{k,t}^{\mathrm{edge}}
\right\}.
\end{equation}
The corresponding energy consumption consists of local computation energy and
transmission energy:
\begin{equation}
\label{eq:energy}
E_{k,t}
=
P_k^{\mathrm{loc}}
D_{k,t}^{\mathrm{loc}}
+
p_{k,t}
\frac{
L_{k,t}\beta_{k,t}
}
{
R_{k,t}10^{6}
},
\end{equation}
where $P_k^{\mathrm{loc}}$ denotes the local computation power consumption.

To characterize the joint impact of computation delay and energy consumption,
we define the effective processing throughput as
\begin{equation}
\label{eq:throughput}
\Phi_{k,t}
=
\frac{
L_{k,t}/10^{6}
}
{
D_{k,t}+\varepsilon
},
\end{equation}
where $\varepsilon$ is a small positive constant for numerical stability.
The corresponding energy efficiency is defined as
$
\eta_{k,t}
=
\frac{
\Phi_{k,t}
}
{
E_{k,t}+\epsilon_E
}.
$

\subsection{Task Definition and User Objective}

A wireless task $\tau$ is characterized by $
\tau=
(\mathcal{H}_{\tau},
\lambda_{\tau},
\bm q_{\tau},
\mathcal C_{\tau},
\bm w_{\tau}),
$
where $\lambda_{\tau}$ denotes the traffic intensity,
$\bm q_{\tau}$ represents QoS requirements,
$\mathcal C_{\tau}$ denotes active constraints, and
$\bm w_{\tau}$ represents objective preference weights. The QoS objective of each task is represented by the task-specific return
$
U_{\tau}
=
F
(
R_{k,t},
D_{k,t},
E_{k,t},
\eta_{k,t};
\bm w_{\tau}
),
$
where larger values indicate better optimization performance. The QoS requirements and active constraints are aggregated into a normalized
violation cost $v_t\geq0$. The environment reward is defined as
\begin{equation}
r_{\mathrm{env}}
=
U_{\tau}
-
\lambda_{\mathrm{pen}} v_t U_{\tau}^{\mathrm{ref}} .
\label{eq:renv}
\end{equation}
where $U_{\tau}$ denotes the task-specific objective,
$v_t$ represents the aggregated normalized QoS and constraint violation,
and $\lambda_{\mathrm{pen}}$ is the penalty coefficient. This reward remains
the semantic anchor of each wireless task, while FARM learns transferable
auxiliary reward guidance across heterogeneous tasks.

\section{Problem Formulation}
\label{problem}
We formulate the considered wireless optimization
problem as a constrained Markov decision process (CMDP). Different from
conventional single-task resource allocation, each wireless task is
characterized by distinct channel configurations, traffic patterns, QoS
requirements, and optimization preferences. Therefore, the objective is not
only to optimize task-specific control policies, but also to learn transferable
reward representations that provide reusable optimization guidance across
heterogeneous wireless tasks.

\subsection{Constrained MDP Formulation}
\label{sec:mdp}

For a given wireless task $\tau$, the interaction between the controller and
the MEC environment is modeled as
$
\mathcal{M}_{\tau}
=
(\mathcal{S},\mathcal{A},P_{\tau},
r_{\mathrm{env}},g_{\tau},\gamma),
$
where $\mathcal{S}$, $\mathcal{A}$, and $P_{\tau}$ denote the state space,
action space, and task-dependent transition probability, respectively.
$r_{\mathrm{env}}$ represents the environment reward determined
by the task objective, $g_{\tau}$ denotes the task-dependent
constraint-cost function, with
$v_t=g_{\tau}(o_t,a_t)$ denoting its normalized instantaneous
violation cost, and $\gamma$ is the discount factor. The system state contains wireless channel information, computation workload,
queue-related states, previous actions, and constraint-related variables. The
joint action is defined as
$
\bm a_t
=
\{(p_{k,t},\beta_{k,t})\}_{k=1}^{N},
$
where $p_{k,t}$ and $\beta_{k,t}$ represent the transmission power and
offloading ratio of user $k$, respectively. The observation and action
dimensions reported in Table~II are defined on a per-user basis: each user
has a 9-dimensional observation vector and a 2-dimensional action vector
$(p_{k,t},\beta_{k,t})$. Accordingly, the joint action consists of $N$
such user-specific action pairs and has dimension $2N$. The transmit-power
allocation satisfies $p_{k,t}\geq 0$ and
$\sum_{k=1}^{N}p_{k,t}\leq P_{\mathrm{tot}}$,
where $P_{\mathrm{tot}}$ denotes the total transmit-power budget. The state
transition is determined by the channel evolution, stochastic task arrivals,
queue dynamics, and computation processes described in Section~\ref{sys}.

\subsection{Task-Dependent Utility and Constraint Modeling}
\label{sec:utility}

Each wireless task $\tau$ is associated with a preference vector
$
\bm w_{\tau}
=
(w^R_{\tau},w^D_{\tau},w^E_{\tau},w^{\eta}_{\tau}),
$
which determines the relative importance of throughput, latency, energy
consumption, and energy efficiency. To construct a unified  objective across heterogeneous tasks, the
individual utility metrics are normalized:
\begin{align}
u^{R}_{k,t}
&=
\sigma(R_{k,t};R^{\min}_{\tau}),
&
u^{D}_{k,t}
&=
\sigma(D^{\max}_{\tau};D_{k,t}),
\nonumber\\
u^{E}_{k,t}
&=
\sigma(E^{\max}_{\tau};E_{k,t}),
&
u^{\eta}_{k,t}
&=
\sigma(\eta_{k,t};\eta^{\min}_{\tau}),
\end{align}
where $
\sigma(x;\nu)=\frac{x}{x+\nu}.
$ The objective utility of user $k$ is defined as $
r^{obj}_{k,t}
=
\bm w_{\tau}^{T}\bm u_{k,t},
$ where $
\bm u_{k,t}
=
(u^R_{k,t},
u^D_{k,t},
u^E_{k,t},
u^\eta_{k,t})^T .
$ The overall objective utility is obtained by averaging over all users:
\begin{equation}
r^{obj}_{t}
=
\frac{1}{N}
\sum_{k=1}^{N}
r^{obj}_{k,t}.
\end{equation}
The normalized objective $r_t^{\mathrm{obj}}$ provides the per-slot
realization of the task-specific objective $U_{\tau}$ used in
\eqref{eq:renv}.
The QoS requirements and active constraints are aggregated into a normalized
violation cost $v_t\geq0$. The SAC controller uses the environment reward
$r_{\mathrm{env}}$ defined in~\eqref{eq:renv}, where
$\lambda_{\mathrm{pen}}$ is the fixed constraint penalty coefficient and
$U_{\tau}^{\mathrm{ref}}$ is the task-specific reference value used for
normalization. This reward preserves the optimization semantics of each
wireless task, while FARM provides transferable auxiliary reward guidance
without replacing the original task objective.

\subsection{Reward-Space Transfer Objective}
\label{sec:multi-task-objective}

The heterogeneous source-task bank is denoted by
$
\mathcal{B}=\{\tau_i\}_{i=1}^{M},
$
where each task $\tau_i$ corresponds to a specific wireless configuration,
objective preference, and constraint setting. Stage~I jointly learns
task-specific SAC controllers and a transferable reward prior through two
coupled optimization processes.
At each iteration, the current ARM generates auxiliary rewards for policy
optimization, while the resulting source-task interaction trajectories are
used to update ARM. The coupling therefore arises through both the auxiliary
reward signal and the trajectory distribution induced by the task-specific
policies.

For each source task $\tau_i$, with the current ARM parameters held fixed
during the policy update, the task-specific controller is optimized as
\begin{align}
\pi_i^{*}
=
\arg\max_{\pi_i}
\mathbb{E}_{\pi_i,\tau_i}
\left[
\sum_{t=0}^{H-1}
\gamma^t
\left(
r_{\mathrm{env},t}
+
\lambda_{\mathrm{aux}}(c_i)
\operatorname{Norm}(\hat r_t)
\right)
\right].
\label{18}
\end{align}
Here, $r_{\mathrm{env},t}$ preserves the task-specific optimization objective,
whereas $\hat r_t$ provides trajectory-conditioned auxiliary guidance generated
by ARM. The policies $\{\pi_i\}$ are optimized only for their corresponding
source tasks and are not transferred to new environments. ARM is updated from the interaction trajectories generated by the current
source-task policies. Let
$\pi_{\Theta}(\hat r_t\mid\Gamma_t,c_i)$ denote the ARM reward actor, which
generates the auxiliary reward $\hat r_t$ from the trajectory window
$\Gamma_t$ and source-task condition $c_i$. Two critics
$Q_{\phi_1}$ and $Q_{\phi_2}$ evaluate the suggested reward together with the
environment state and action. During an ARM update, the critic parameters are
first optimized by minimizing $\mathcal{L}_{Q}$, followed by the reward-actor
parameters through $\mathcal{L}_{\pi}$:
$
\phi^{*}
=
\arg\min_{\phi}
\mathcal{L}_{Q}(\phi),
$$
\Theta^{*}
=
\arg\min_{\Theta}
\mathcal{L}_{\pi}(\Theta),
$$
\mathcal{L}_{\mathrm{reward}}(\Theta,\phi)
=
\mathcal{L}_{Q}(\phi)
+
\mathcal{L}_{\pi}(\Theta),
$
where $\phi=\{\phi_1,\phi_2\}$ denotes the parameters of the twin ARM critics.
Here, $\mathcal{L}_{\mathrm{reward}}$ provides compact notation for the two
ARM optimization losses and does not imply a single joint gradient update.
The dependence of each loss on the current source-task policies and the
counterpart ARM parameters is omitted from the notation for clarity, since
these quantities are held fixed during the corresponding update. For a replay transition at slot $t$, the soft Bellman target is defined as
\begin{equation}
\begin{aligned}
y_t
={}&
r_{\mathrm{env},t}
+
(1-d_t)\gamma
\Big[
\min_{j\in\{1,2\}}
Q_{\bar{\phi}_j}
\big(
o_{t+1},a_{t+1},\hat r_{t+1}
\big)
\\
&\hspace{19mm}
-
\kappa_{\mathrm{ARM}}
\log
\pi_{\Theta}
\big(
\hat r_{t+1}\mid\Gamma_{t+1},c_i
\big)
\Big],
\end{aligned}
\label{eq:arm_target}
\end{equation}
where
$\hat r_{t+1}\sim
\pi_{\Theta}(\cdot\mid\Gamma_{t+1},c_i)$,
$a_{t+1}$ is generated by the corresponding Stage-I task policy,
$d_t\in\{0,1\}$ is the terminal indicator,
$\bar{\phi}_j$ denotes the Polyak target-critic parameters, and
$\kappa_{\mathrm{ARM}}$ is the ARM entropy coefficient. The twin critics are trained by minimizing
\begin{equation}
\mathcal{L}_{Q}(\phi)
=
\sum_{j=1}^{2}
\mathbb{E}
\left[
\left(
Q_{\phi_j}(o_t,a_t,\hat r_t)
-
y_t
\right)^2
\right],
\label{eq:arm_critic_loss}
\end{equation}
where $\hat r_t$ is detached from the reward actor when updating the critics.
The reward actor is subsequently optimized using
\begin{equation}
\mathcal{L}_{\pi}(\Theta)
=
\mathbb{E}
\left[\!
\kappa_{\mathrm{ARM}}
\log
\pi_{\Theta}
\big(\!
\hat r_t\mid\Gamma_t,c_i
\!\big)
\!\!
-\!\!
\min_{j\in\{1,2\}}\!
Q_{\phi_j}\!(o_t,a_t,\hat r_t)
\!\right]\!,
\label{eq:arm_actor_loss}
\end{equation}
with
$\hat r_t\sim\pi_{\Theta}(\cdot\mid\Gamma_t,c_i)$.
The target critics are then updated through Polyak averaging. Thus, Stage~I alternates between task-specific policy optimization and ARM
reward-prior learning. The SAC controllers optimize their corresponding
wireless tasks using the environment reward augmented by ARM guidance, while
ARM learns transferable trajectory-level reward knowledge from the interaction
distributions induced by these controllers. The Bellman target remains
anchored to the original environment reward $r_{\mathrm{env}}$, so the learned
reward prior provides auxiliary guidance without directly imitating or
replacing the task-specific environment reward.

After Stage~I, the learned reward prior is transferred to an unseen target
task $\tau_{\mathrm{tar}}$. The pretrained ARM backbone is retained as the
transferable reward representation, while target-specific adaptation is
performed for the new environment. Let $c_{\mathrm{tar}}$ denote the condition
vector of $\tau_{\mathrm{tar}}$. At the policy level, the target controller is
optimized as
\begin{align}
\max_{\pi_{\mathrm{tar}}}
&
\mathbb E_{\pi_{\mathrm{tar}},\tau_{\mathrm{tar}}}
\left[
\sum_{t=0}^{H-1}
\gamma^t
\left(
r_{\mathrm{env},t}
+
\lambda_{\mathrm{aux}}(c_{\mathrm{tar}})
\operatorname{Norm}(\hat r_t)
\right)
\right]
\nonumber\\
\mathrm{s.t.}\quad
&
\mathbb E_{\pi_{\mathrm{tar}},\tau_{\mathrm{tar}}}
\left[
\sum_{t=0}^{H-1}
\gamma^t v_t
\right]
\leq d_{\mathrm{tar}} .
\label{eq:target-transfer}
\end{align}
where $d_{\mathrm{tar}}$ denotes the admissible discounted
constraint-violation budget for target task $\tau_{\mathrm{tar}}$.
The pretrained ARM backbone remains frozen during Stage-II transfer, while
the lightweight target-specific adaptation used by objective-routed adapter fine-tuning (ObjRoute AFT) is detailed
in Section~\ref{sec:alg-stage2}. This formulation separates transferable
reward knowledge from task-specific policy optimization: ARM captures
reusable trajectory-level evaluation knowledge across heterogeneous source
tasks, whereas individual controllers remain specialized to their respective
wireless environments.
\section{Algorithm Design of FARM}
\label{algorithm}

\begin{figure}[!t]
\centering
\includegraphics[width=0.45\textwidth]{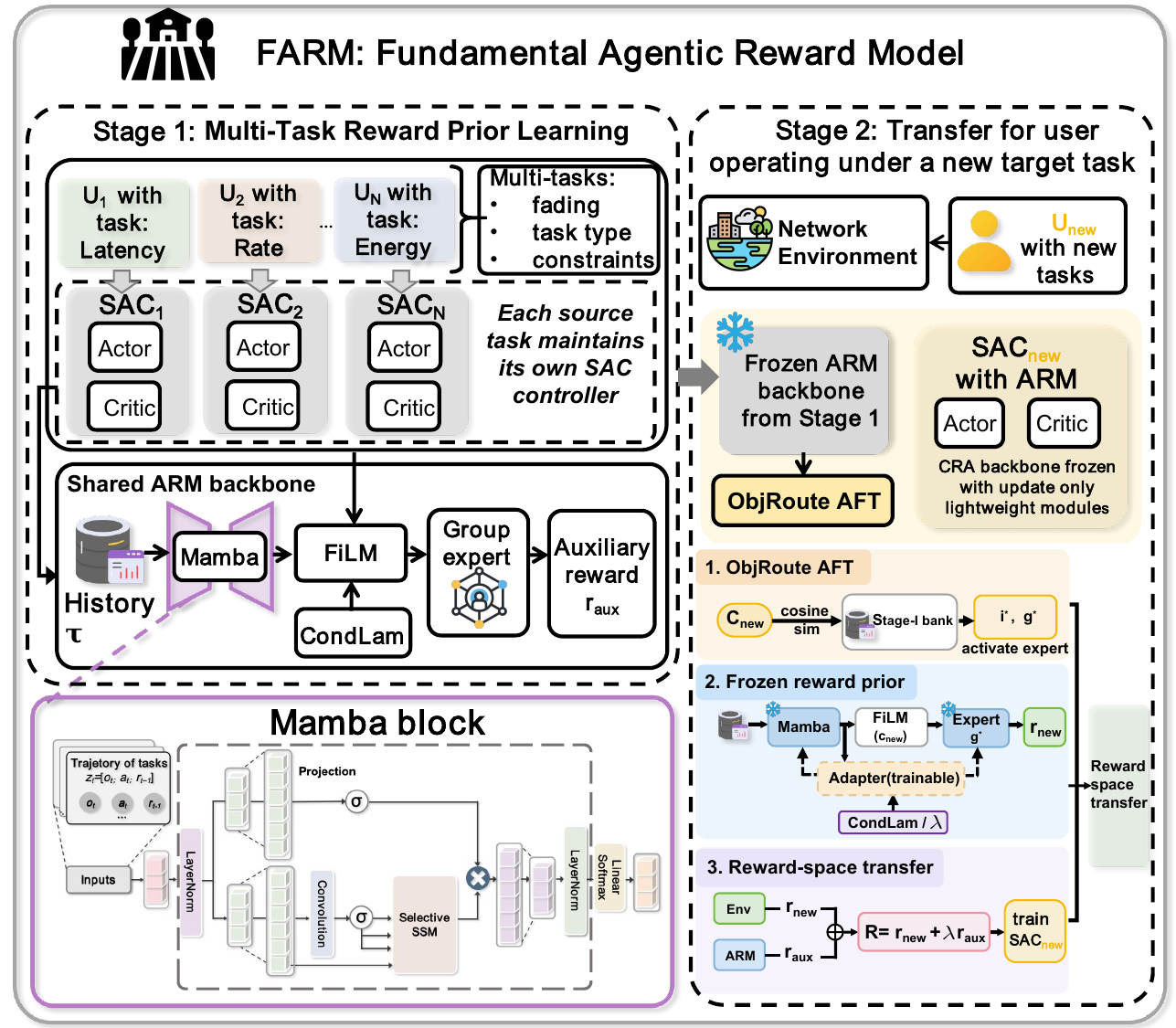}
\caption{
Overall architecture of FARM. Stage-I learns a transferable trajectory-level reward prior from heterogeneous wireless tasks using ARM. In Stage-II, the pretrained ARM backbone is frozen, while a lightweight target-specific adapter and the target SAC controller are optimized.
}
\label{fig:farm_framework}
\end{figure}

FARM decouples transferable reward knowledge from task-specific policy
optimization by learning a trajectory-level reward prior from heterogeneous tasks.
As illustrated in Fig.~\ref{fig:farm_framework}, FARM follows a two-stage
Reward-Space Transfer framework. In Stage~I, ARM learns a reward prior
from a heterogeneous source-task bank, while individual SAC controllers retain
task-specific policy optimization. Within ARM, the Mamba trajectory encoder
captures temporal dependencies, and task conditioning with objective-grouped
reward modeling extracts reusable evaluation knowledge across tasks. In
Stage~II, the pretrained reward prior is frozen and reused through ObjRoute AFT to guide a target-specific SAC controller for an unseen task without
directly inheriting a source policy. The two-stage procedure is
summarized in Algorithm~\ref{alg:farm}.

\subsection{Design Principle: Agentic Reward Modeling}
\label{sec:alg-principle}

Let $r_{\mathrm{env}}$ denote the environment reward
in~\eqref{eq:renv}, which reflects the task objective
and QoS constraints in the wireless environment. FARM does not replace the original optimization
objective with a learned reward representation but generates an auxiliary reward $\hat r_t$ by reasoning over
historical trajectories and task conditions. The training reward is formulated
as
$
r^{\mathrm{train}}_t
=
r_{\mathrm{env}}
+
\lambda_{\mathrm{aux}}(c)
\mathrm{Norm}(\hat r_t),
$
where $\mathrm{Norm}(\cdot)$ denotes online normalization with optional
clipping. This formulation preserves the optimization objective,
allowing the transferable reward prior learned by ARM to provide additional guidance for temporal credit assignment and policy optimization.

\subsection{Task Condition Interface}
\label{sec:alg-condition}

Heterogeneous MEC tasks differ in channel configurations, objective
preferences, and constraint requirements. Therefore, FARM represents each task
through a condition vector
$
c
=
\mathrm{EncodeTask}
(
\mathcal{H}_{\tau},
\bm w_{\tau},
\bm q_{\tau},
\mathcal C_{\tau},
\mathrm{sev}_{\tau}
),
$
where $\mathcal{H}_{\tau}$ represents the channel configuration,
$\bm w_{\tau}$ denotes objective preferences, $\bm q_{\tau}$ represents QoS
requirements, and $\mathcal C_{\tau}$ denotes active constraints.
The severity descriptor $\mathrm{sev}_{\tau}$ summarizes channel severity,
traffic intensity, and resource limitations. The mapping $\mathrm{EncodeTask}(\cdot)$ is deterministic and produces a
$d_c$-dimensional task-condition vector. The condition vector remains constant
for a given task and does not depend on slot-wise observations. The ARM reward encoder, FiLM module, objective-specific
experts, and CondLam module all take $c$ as input, enabling
both reward generation and auxiliary-reward weighting to adapt
to heterogeneous task conditions.

\subsection{Agentic Reward Model for Transferable Reward Reasoning}
\label{sec:alg-arm}

\begin{itemize}

\item
\textbf{Trajectory representation:} At each time slot, ARM receives historical interaction information to extract
trajectory-level evaluation knowledge. The input token is defined as
$
z_t=[o_t;a_t;r_{t-1}],
$ where $o_t$ represents the system observation, $a_t$ denotes the MEC action,
and $r_{t-1}$ is the previous reward. The trajectory window is defined as $
\Gamma_t
=
(z_{t-W+1},...,z_t).
$

\item
\textbf{Trajectory encoder:} The trajectory encoder extracts temporal reward-related representations:
$
\psi_t
=
\mathrm{Enc}_{\Theta}(\Gamma_t),
$ where $\psi_t$ denotes the reward representation generated by ARM. The encoder
can be implemented using either Mamba or a causal Transformer while maintaining
the same trajectory interface.

\item
\textbf{Condition-aware modulation:} The task condition is integrated through FiLM:
$
\psi'_t
=
\gamma(c)\odot\psi_t+\beta(c),
$ where $\gamma(c)$ and $\beta(c)$ are condition-dependent modulation
parameters.

\item
\textbf{Objective-specific experts:} To capture different optimization preferences across heterogeneous wireless
tasks, FARM adopts grouped reward experts $
\{\mathrm{Expert}_{g}\}_{g\in\mathcal G}.
$ The auxiliary reward is given as
$
\hat r_t
=
s
\tanh
(
\mathrm{Expert}_{G(c)}(\psi'_t)
),
$ where $G(c)$ selects the objective expert and $s$ controls the reward scale.

\item
\textbf{Adaptive reward scaling:} The contribution of the auxiliary reward is determined by
$
\lambda_{\mathrm{aux}}(c)
=
\mathrm{CondLam}(c),
$ allowing the shaping strength to adapt to the task condition.
The $0.3$ in Table~II denotes the base scale
of CondLam.
The realized $\lambda_{\mathrm{aux}}(c)$ varies with the task
condition and is clipped to $[0.05,1.0]$ for stable policy
optimization.

\end{itemize}

\subsection{Stage~I: Multi-Task Reward Prior Learning}
\label{sec:alg-stage1}

Stage~I aims to learn a transferable reward prior from a heterogeneous wireless
task bank $\mathcal B$. The objective is to extract shared trajectory
evaluation knowledge across diverse wireless tasks while preserving the
task-specific optimization capability of individual controllers. The transferable reward prior parameters are defined as
$
\Theta=
\{
\mathrm{Enc},
\mathrm{FiLM},
\{\mathrm{Expert}_{g}\},
\mathrm{CondLam}
\}.
$

During training, each source task $\tau_i$ maintains an independent SAC policy
$\pi_i$, while ARM is jointly optimized to capture reusable reward knowledge
across heterogeneous tasks. Different from policy sharing approaches, FARM
does not enforce a common controller among source tasks. Instead, the learned
reward prior serves as a transferable knowledge interface that can be reused
for unseen tasks. The learned ARM captures transferable trajectory evaluation patterns across
heterogeneous wireless tasks, while the task-specific SAC policies only serve
as optimization agents for their corresponding source environments. Therefore,
Stage~I separates transferable reward knowledge learning from task-dependent
control policy optimization, enabling the learned reward prior to be reused
across different wireless task distributions.

\begin{algorithm}[!t]
\caption{FARM Reward Prior Learning and Transfer}
\label{alg:farm}

\begin{algorithmic}[1]

\Statex \textbf{Stage I: Multi-Task Reward Prior Learning}

\Require
Task bank
$\mathcal{B}=\{\tau_i\}_{i=1}^{M}$;
window length $W$; training steps $T_{\mathrm{S1}}$

\Ensure
Transferable reward prior $\Theta$;
task-specific policies $\{\pi_i\}$

\State Initialize reward-prior parameters $\Theta$, twin ARM critic
parameters $\phi=\{\phi_1,\phi_2\}$, and $\{\pi_i\}$

\For{$t=1,\ldots,T_{\mathrm{S1}}$}
    \For{each $\tau_i\in\mathcal{B}$}
        \State Execute $a_t\sim\pi_i(\cdot|o_t)$ and observe
        $r_{\mathrm{env}}$
        \State Construct trajectory window $\Gamma_t$
        \State Generate auxiliary reward
        $\hat r_t=f_{\Theta}(\Gamma_t,c_i)$
        \State Update $\pi_i$ using SAC with
        $r_{\mathrm{env}}
        +\lambda_{\mathrm{aux}}(c_i)\mathrm{Norm}(\hat r_t)$
        \State Update ARM parameters $(\Theta,\phi)$ using
$\mathcal{L}_{\mathrm{reward}}(\Theta,\phi)$
    \EndFor
\EndFor

\State Store the learned reward prior $\Theta$

\vspace{1mm}

\Statex \textbf{Stage II: Reward-Space Transfer}

\Require
Pretrained reward prior $\Theta$;
target condition $c_{\mathrm{tar}}$;
adaptation steps $K_{\mathrm{tar}}$

\Ensure
Adapted target policy $\pi_{\mathrm{tar}}$

\State Freeze reward-prior parameters $\Theta$
\State Initialize target-specific adapter parameters
$\omega_{\mathrm{tar}}$
\State Select objective prototype
$
i^\star=\arg\max_i \mathrm{cos}(c_{\mathrm{tar}},c_i)
$
\State Activate objective expert
$g^\star=G(c_{i^\star})$
\State Initialize target policy $\pi_{\mathrm{tar}}$

\For{$n=1,\ldots,K_{\mathrm{tar}}$}
    \State Collect target trajectory and construct $\Gamma_t$
    \State Generate auxiliary reward
    $\hat r_t=f_{\Theta,\omega_{\mathrm{tar}}}
    (\Gamma_t,c_{\mathrm{tar}})$
    \State Update target adapter $\omega_{\mathrm{tar}}$
    through ObjRoute AFT
    \State Update $\pi_{\mathrm{tar}}$ using SAC with
    $r_{\mathrm{env}}
    +\lambda_{\mathrm{aux}}(c_{\mathrm{tar}})
    \mathrm{Norm}(\hat r_t)$
\EndFor

\State \Return $\Theta,\{\pi_i\},\pi_{\mathrm{tar}}$

\end{algorithmic}
\end{algorithm}

\subsection{Stage~II: Reward-Space Transfer}
\label{sec:alg-stage2}

After Stage~I, FARM transfers the learned reward prior to an unseen target
task $\upsilon$. FARM
transfers trajectory-level reward knowledge while optimizing a target-specific
SAC controller for the new environment. During Stage~II, the pretrained ARM
backbone $\Theta$ is frozen, while a lightweight target-specific adapter with
parameters $\omega_{\upsilon}$ is introduced for ObjRoute AFT. Within the
reward-prior pathway, only $\omega_{\upsilon}$ is optimized, leaving the
pretrained ARM parameters unchanged; the target SAC policy
$\pi_{\upsilon}$ is optimized separately for the new environment.
The frozen ARM backbone together with the target-specific adapter provides
auxiliary reward guidance conditioned on the target-task vector
$c_{\upsilon}$. FARM first employs objective-aware routing to select the most relevant
source prototype according to the cosine similarity between task-condition
vectors,
\begin{equation}
\label{eq:objroute}
i^\star
=
\arg\max_i
\mathrm{cos}(c_\upsilon,c_i)
=
\arg\max_i
\frac{c_\upsilon^{\!\top}c_i}
{\|c_\upsilon\|_2\,\|c_i\|_2}.
\end{equation}

The corresponding objective expert $g^\star$ is then activated
based on the selected prototype, allowing ARM to generate
reward guidance consistent with the target optimization
preference. Although the pretrained reward prior is frozen
during adaptation, its outputs remain conditioned on the
target-task vector $c_\upsilon$. In particular, CondLam evaluates
$c_\upsilon$ to produce the target-dependent auxiliary weight
$\lambda_{\mathrm{aux}}(c_\upsilon)$, while the condition-aware
reward pathway generates $\hat{r}_t$. The target policy
$\pi_\upsilon$ is therefore optimized using the environment
reward together with the resulting transferred auxiliary
guidance. In this way, FARM performs Reward-Space Transfer
without directly transferring source-task policies.














\section{Experimental Results}
\label{num}

This section evaluates FARM for heterogeneous MEC optimization through three
research questions: \textbf{RQ1: Multi-task reward prior learning.}
Can FARM learn reusable trajectory-level reward knowledge across
heterogeneous source tasks while preserving task-specific policy optimization,
compared with independent SAC, direct policy sharing, and conventional shared
reward modeling? \textbf{RQ2: Reward prior transfer to unseen tasks.}
Can the learned reward prior provide effective adaptation guidance for
previously unseen tasks under shifts in channel conditions, traffic patterns,
QoS requirements, and optimization objectives, compared with training from
scratch and conventional reward transfer? \textbf{RQ3: Impact of trajectory modeling.}
How does the temporal representation backbone affect Reward-Space Transfer,
particularly when the history length exceeds that used during Stage-I
pretraining?

The experiments are implemented on a Linux server running Ubuntu~24.04 with
dual-socket AMD EPYC~7713 CPUs and eight NVIDIA A100-SXM4 80\,GB GPUs.
The software environment includes Python~3.12, PyTorch~2.11 with CUDA~12.8,
and mamba-ssm~2.3.2. All experiments use the MEC model described in
Section~\ref{sys} and the optimization formulation in
Section~\ref{problem}. The experimental configuration and compared methods are summarized in
Tables~\ref{tab:exp-setup} and~\ref{tab:exp-methods}, respectively. The compared methods
span independent task optimization, direct policy sharing, conventional
shared-reward learning, and the proposed reward-prior transfer framework;
the Transformer variant differs from FARM only in the trajectory encoder and
therefore serves as a controlled backbone ablation.

\textbf{Common Experimental Setup}
All experiments follow the heterogeneous MEC environment and optimization
model defined in Sections~\ref{sys} and~\ref{problem}. Task heterogeneity
is induced by variations in channel statistics, temporal correlations,
traffic conditions, QoS requirements, constraints, and optimization
objectives. All compared methods use the same environment reward
in~\eqref{eq:renv} and are evaluated using the task-specific objective
$U_{\tau}$, with larger values indicating better performance. Since different
objective families have distinct physical meanings and numerical scales,
absolute objective values are reported within each family, while relative
performance gains are computed only under matched task and evaluation
settings.

\subsection{Stage-I and Stage-II Task Protocol}
\label{sec:exp-tasks}

\paragraph{Stage-I source task bank}
Stage~I contains $M=10$ MEC tasks spanning Rate, Latency, Energy, and
balanced Rate--Latency objectives. The tasks differ in fading distributions,
temporal correlations, traffic conditions, QoS thresholds, and constraint
configurations, thereby forming a heterogeneous source bank for reward-prior
learning. Each source task maintains its own SAC controller, while ARM learns
shared trajectory-level reward knowledge across tasks. Unless otherwise
specified, FARM uses a history window of $W=32$ with objective-grouped
experts, FiLM conditioning, and CondLam. Stage-I evaluation is designed to distinguish three forms of knowledge reuse:
independent task optimization, direct policy sharing, and shared reward
modeling. This setting therefore examines whether transferable reward
knowledge can be learned without forcing heterogeneous tasks to share a common
action-generation policy.

\paragraph{Stage-II unseen tasks}
Stage~II evaluates transfer to held-out users whose task configurations are
not jointly observed during Stage-I training. The target set contains Energy,
Latency, Rate--Energy, and Rate--Latency objectives with different channel
conditions, QoS requirements, and difficulty levels, representing increasing
degrees of distribution shift from the source-task bank. During transfer, the Stage-I reward prior remains frozen. No source-task policy
is transferred to the target environment; instead, a target-specific SAC
controller is optimized using the original environment reward together with
the auxiliary reward generated from the frozen prior through ObjRoute AFT. Although the Stage-I reward-prior parameters remain frozen,
the prior is still evaluated with each target-task condition
$c_\upsilon$. Consequently, both the generated auxiliary reward
$\hat{r}_t$ and its CondLam weight
$\lambda_{\mathrm{aux}}(c_\upsilon)$ remain target dependent
during Stage-II adaptation.
Unless otherwise specified, Stage-II adaptation uses
$40\times10^{3}$ environment steps. This protocol evaluates whether reward
knowledge learned from heterogeneous source tasks remains useful for policy
optimization under previously unseen task configurations.

\paragraph{History-length evaluation}
RQ3 isolates the effect of temporal trajectory representation. Both the
Mamba-$L{=}2$ and Transformer-$L{=}2$ reward priors are pretrained with
$W=32$ and then frozen before Stage-II adaptation. At inference time, they are
evaluated with $W\in\{32,64,128\}$ on held-out users, without
Stage-I retraining. This setting tests whether the learned temporal
representations remain effective when longer interaction histories are
introduced and separates temporal-backbone effects from the remaining FARM
components.

\begin{table}[!t]
\centering
\footnotesize
\setlength{\tabcolsep}{3.3pt}
\renewcommand{\arraystretch}{0.97}
\caption{Experimental configuration.}
\label{tab:exp-setup}

\begin{tabular}{@{}p{0.4\linewidth}p{0.59\linewidth}@{}}
\toprule
Parameter & Setting \\
\midrule

\multicolumn{2}{@{}l}{\textit{Wireless environment}} \\

Bandwidth / noise
& $10\,\mathrm{MHz}$ / $0.5$ (normalized AWGN) \\

Local / edge CPU
& $1\,\mathrm{GHz}$ / $2\,\mathrm{GHz}$ \\

Total transmit-power budget
& $P_{\mathrm{tot}}=8$ W \\

Power normalization reference
& $10$ W \\

Payload / CPU intensity
& $10^{6}$ bits / $10^{3}$ cycles/bit \\

Channel model
& Rayleigh, Rician, Nakagami-$m$;
task-wise AR coefficient $\rho_{\tau}$ \\

Episode horizon
& $H=200$ slots \\

Constraint penalty coefficient
& $\lambda_{\mathrm{pen}}=1.5$ \\

\midrule
\multicolumn{2}{@{}l}{\textit{SAC controller}} \\

Per-user observation/action interface
& obs-dim $=9$, act-dim $=2$ $(p,\beta)$ \\

Network
& MLP, hidden dim. $256$, two residual blocks \\

Optimization
& Adam, lr $=3\times10^{-4}$,
$\gamma=0.99$, Polyak $=0.005$ \\

Replay
& Batch $256$, buffer $10^6$ \\

Entropy
& Automatic temperature tuning; target entropy
$=-\mathrm{act\text{-}dim}$ \\

\midrule
\multicolumn{2}{@{}l}{\textit{Reward prior and evaluation}} \\

ARM representation
& $W=32$, $d_{\mathrm{model}}=256$;
RQ3: $W\in\{32,64,128\}$ \\

ARM optimization
& Batch $128$, lr $=3\times10^{-5}$ \\

Conditioning
& FiLM + objective experts + CondLam MLP \\

Conditional auxiliary weighting
& $\lambda_{\mathrm{aux}}(c)=\operatorname{CondLam}(c)$;
base scale $0.3$, output clipped to $[0.05,1.0]$ \\

Stage-II transfer
& Frozen reward prior + ObjRoute AFT \\

Training horizon
& Stage~I/II: $40$k steps; RQ3: $30$k steps \\

Evaluation
& Every $500$ steps; final $15\%$ average \\

\bottomrule
\end{tabular}
\end{table}

\begin{table}[!t]
\centering
\small
\footnotesize
\setlength{\tabcolsep}{2.5pt}
\renewcommand{\arraystretch}{0.97}
\caption{Compared methods. All methods use the reward in
\eqref{eq:renv} and the same SAC interface unless otherwise specified.}
\label{tab:exp-methods}

\begin{tabular}{@{}p{0.25\linewidth}p{0.12\linewidth}p{0.57\linewidth}@{}}
\toprule
Method & Stage & Description \\
\midrule

Single-task SAC
& I, II
& Independent SAC without cross-task transfer; scratch SAC is used for
unseen targets. \\

Shared MTSAC
& I
& One actor--critic trained from mixed Stage-I replay to evaluate direct
policy sharing. \\

CRA
& I
& Shared auxiliary-reward baseline without ARM task conditioning or
trajectory-level reward reasoning. \\

CRA Transfer
& II
& Frozen CRA reward model reused to guide target SAC adaptation. \\

FARM
& I$\rightarrow$II
& ARM with Mamba ($L\in\{2,4\}$); the learned reward prior is frozen and
reused through ObjRoute AFT. \\

Transformer
& I$\rightarrow$II
& Matched FARM ablation replacing Mamba with a causal Transformer
($L\in\{2,4\}$); all other components are unchanged. \\

\bottomrule
\end{tabular}

\vspace{0.5mm}
\raggedright
\scriptsize
\textit{Note:} Stage~I uses $M=10$ heterogeneous source tasks.
Transformer is a backbone ablation of FARM.
\end{table}

\subsection{Evaluation Metrics}
\label{sec:exp-report}

We consider three research questions. For \textbf{RQ1}, Stage-I performance is assessed using task-specific objective values $U_{\tau}$, convergence behavior, and late-stage performance relative to independently trained Single-task SAC~\cite{haarnoja2018soft} . Because the objective families have different physical meanings and scales, task-wise results and within-family aggregates are reported instead of directly averaging absolute objective values across all source tasks. Comparisons with Shared MTSAC and CRA~\cite{NEURIPS2025_eeb273b2} further distinguish the effects of direct policy sharing and conventional shared reward modeling from FARM's task-conditioned reward-prior learning.

For \textbf{RQ2}, transfer performance is evaluated using both adaptation
trajectories and late-stage objective values on held-out users. To normalize
the different objective scales, we report the relative late-stage gain over a
SAC controller trained from scratch on the same target task,
$
\Delta_{\mathrm{SAC}}
=
100
\frac{U_{\mathrm{method}}-U_{\mathrm{SAC}}}
{|U_{\mathrm{SAC}}|}
\,[\%],
$
where a positive value indicates beneficial transfer and a larger value corresponds to a greater improvement over scratch SAC. 

For \textbf{RQ3}, Mamba-$L{=}2$ and Transformer-$L{=}2$ are compared under
matched Stage-I and Stage-II configurations. Both reward priors are pretrained
with $W=32$ and remain frozen during transfer. Evaluation on held-out users
$u_1$, $u_3$, and $u_8$ with $W\in\{32,64,128\}$ reports both the relative
late-stage gain of Mamba over Transformer and the corresponding absolute
objective gap. This dual representation avoids over-interpreting relative
percentage changes when the absolute returns are close to zero and directly
examines the robustness of the two trajectory encoders as the history length
increases.

All comparisons use matched evaluation horizons, environment
configurations, and interaction budgets within each experiment.
Late-stage metrics average the final $15\%$ of logged evaluations
within each run (Table~\ref{tab:exp-setup}).





\subsection{Stage-I Multi-Task Reward Prior Learning}
\label{sec:exp-rq1}

We first address \textbf{RQ1} by examining whether ARM can learn transferable
reward knowledge across heterogeneous source tasks while preserving
task-specific policy optimization. The Stage-I task bank contains $M=10$ MEC
tasks spanning different objectives, fading characteristics, temporal
correlations, traffic conditions, and QoS constraints. Following the common
configuration in Table~\ref{tab:exp-setup}, all methods use the same
environment reward, observation--action interface, and interaction budget. We compare four learning paradigms: Single-task SAC, which independently
optimizes one controller for each task; Shared MTSAC, which trains a single
actor--critic from mixed multi-task experience; CRA, which introduces a shared
auxiliary-reward model; and FARM, which retains task-specific SAC controllers
while sharing a trajectory-level reward prior through ARM. The temporal
backbone comparison is examined separately in \textbf{RQ3}.

\begin{figure*}[!t]
\centering
\includegraphics[width=0.88\textwidth]
{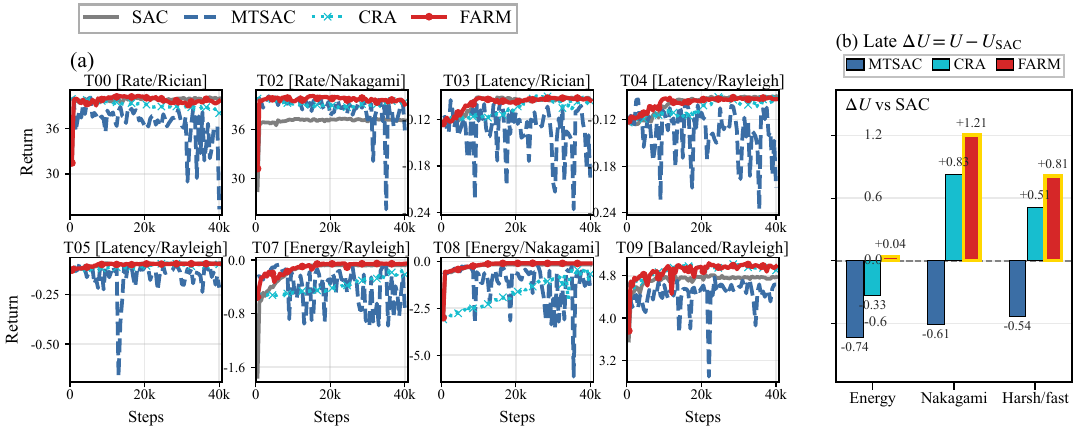}
\caption{
Stage-I reward-prior learning across heterogeneous source tasks.
\textbf{(a)} Representative learning curves for Single-task SAC, Shared MTSAC,
CRA, and FARM; higher $U_{\tau}$ indicates better task-specific performance.
\textbf{(b)} Mean late-stage objective gap relative to Single-task SAC,
$\Delta U=U_{\mathrm{method}}-U_{\mathrm{SAC}}$, for representative
challenging task groups. Positive values indicate improvement over
Single-task SAC.
}
\label{fig:s1-bank}
\end{figure*}

Fig.~\ref{fig:s1-bank}(a) shows the Stage-I learning trajectories across
representative Rate, Latency, Energy, and balanced Rate--Latency tasks.
Because the corresponding objective values have different numerical scales,
their raw returns are interpreted within each task. To summarize the behavior on challenging task groups without introducing
instability from near-zero SAC returns, Fig.~\ref{fig:s1-bank}(b) reports the
absolute late-stage gap
$
\Delta U
=
U_{\mathrm{method}}-U_{\mathrm{SAC}},
$
averaged over the tasks in each group. Unlike a relative percentage metric,
this difference does not divide by $|U_{\mathrm{SAC}}|$ and therefore avoids
artificially inflated ratios when the SAC reference value is close to zero.
The group averages are used as descriptive summaries of the late-stage
performance difference. The results first reveal the limitation of direct policy sharing. Shared MTSAC
remains below Single-task SAC in all three challenging groups, with
$\Delta U=-0.737$, $-0.608$, and $-0.537$ for Energy, Nakagami, and
harsh/fast-fading tasks, respectively. This behavior is consistent with the
learning curves in Fig.~\ref{fig:s1-bank}(a), where a single shared
actor--critic exhibits pronounced degradation on several heterogeneous source
tasks. The result indicates that directly sharing action-generation parameters
can introduce cross-task interference when objective preferences, channel
dynamics, and constraints differ substantially. CRA alleviates part of this degradation but remains sensitive to task
heterogeneity. Its late-stage gap is $-0.331$ on the Energy group, whereas it
becomes positive on Nakagami and harsh/fast-fading tasks, reaching
$+0.831$ and $+0.512$, respectively. Thus, shared auxiliary-reward guidance
can improve multi-task learning relative to direct policy sharing, but its
effectiveness is not uniform across different task conditions. FARM provides the most consistent behavior across the three groups. It remains
slightly above Single-task SAC on Energy tasks with $\Delta U=+0.041$, while
achieving substantially larger gains of $+1.207$ on Nakagami tasks and
$+0.812$ under harsh/fast-fading conditions. Importantly, the Energy result
shows that FARM does not obtain an artificially large advantage when the SAC
reference return is close to zero; instead, the absolute improvement is
modest but positive. The larger gains under Nakagami and harsh/fast-fading
conditions indicate that the learned reward prior is particularly useful when
channel variability and task heterogeneity make direct policy sharing more
difficult.

Overall, the Stage-I results support the main design principle of FARM.
cross-task knowledge is shared through a task-conditioned trajectory-level
reward prior, while action generation remains task specific. Compared with
Shared MTSAC, this separation reduces cross-task interference, and compared
with CRA, it provides more consistent reward guidance across the evaluated
challenging conditions. The influence of the temporal trajectory encoder is
isolated separately in \textbf{RQ3}.

\subsection{Stage-II Transfer to Seen and Unseen Tasks}
\label{sec:exp-rq2}

We next evaluate \textbf{RQ2}: whether the reward prior learned in Stage~I
can be reused to guide target-policy adaptation under previously unseen task
conditions. Unlike Stage~I, which learns reward knowledge from the
heterogeneous source-task bank, Stage~II freezes the pretrained reward prior
and optimizes a target-specific SAC controller through ObjRoute AFT
(Algorithm~\ref{alg:farm}). No source-task policy is directly
transferred. All Stage-II experiments use $40\times10^{3}$ interaction steps with
$\mathrm{learn\_start}=0$ and $W=32$. We consider three levels of task
familiarity and distribution shift. The \textbf{seen (ID)} setting reuses
Stage-I tasks T02 (Nakagami Rate) and T09 (balanced Rate--Latency) to examine
whether previously learned reward knowledge remains useful when adapting a new
target controller. The \textbf{moderate-OOD} setting uses FAR-M, whose fading
characteristics, dual QoS requirements, and power constraints form a task
configuration absent from the Stage-I bank. The \textbf{hard open-set}
setting consists of users $u_{15}$ and $u_{23}$, representing challenging
Rate--Latency tasks under Rayleigh and Nakagami fading, respectively.

\begin{figure*}[!t]
\centering
\includegraphics[width=0.88\textwidth]{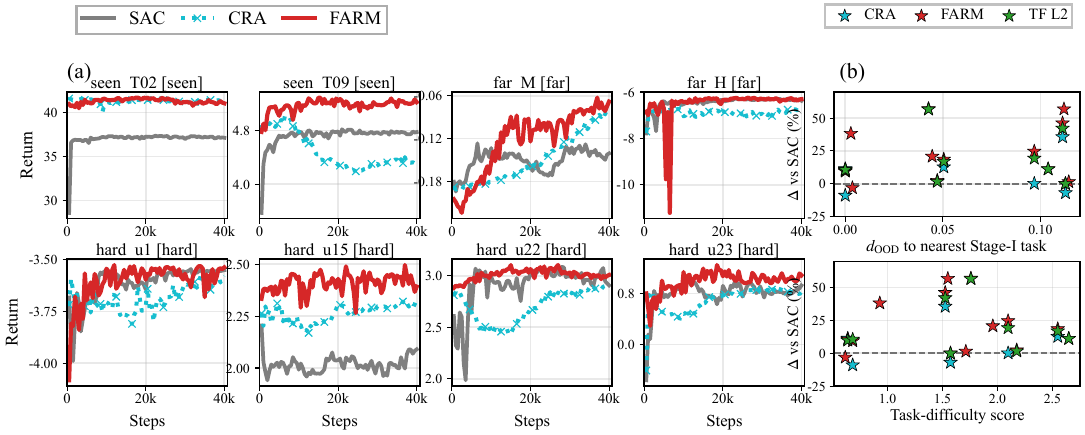}
\caption{
Stage-II Reward-Space Transfer under heterogeneous target-task shifts.
\textbf{(a)} Adaptation trajectories using the frozen FARM reward prior and
ObjRoute AFT on seen, moderate-OOD, and hard open-set users; higher
$U_{\tau}$ indicates better performance.
\textbf{(b)} Late-stage gain over scratch SAC versus the distance to the
nearest Stage-I task and target-task difficulty.
}
\label{fig:stage2-transfer}
\end{figure*}

\begin{figure}[!t]
\centering
\includegraphics[width=0.6\linewidth]
{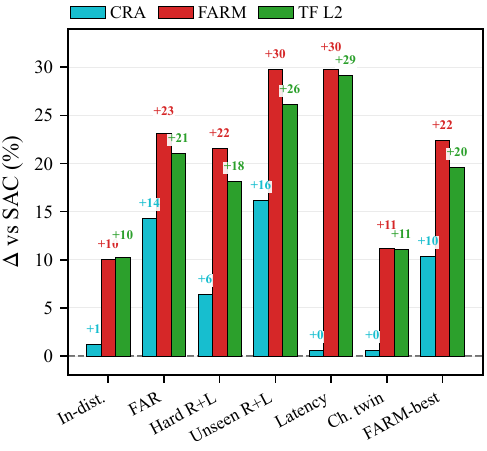}
\caption{
Stage-II transfer gain aggregated by user family.
Bars report the mean late-stage improvement over SAC trained from scratch for
CRA Transfer, FARM (Mamba-$L{=}2$), and the matched
Transformer-$L{=}2$ variant. 
}
\label{fig:stage2-family}
\end{figure}

Fig.~\ref{fig:stage2-transfer}(a) shows the Stage-II adaptation trajectories, while
Table~\ref{tab:s2-agg} reports the corresponding late-stage gain
$\Delta_{\mathrm{SAC}}$ over SAC trained from scratch under the matched
target-task protocol. CRA Transfer and FARM use the same Stage-II interaction
budget, allowing the contribution of transferable reward guidance to be
compared directly. For the seen tasks, FARM achieves a $+10.0\%$ improvement over Single-task
SAC, compared with only $+1.2\%$ for CRA Transfer. The matched
Transformer-$L{=}2$ variant reaches $+10.2\%$, which is nearly identical to
the Mamba-based FARM result. This indicates that, when the target task remains
close to the Stage-I distribution, the main benefit comes from reusing the
learned reward prior instead of from the choice of temporal backbone.

The benefit of Reward-Space Transfer becomes more pronounced under unseen task configurations. 
For the combined FAR-M, $u_{15}$, and $u_{23}$ group, FARM achieves a mean gain of $+29.8\%$, compared with $+26.1\%$ for the Transformer variant and $+16.1\%$ for CRA Transfer. 
In the moderate-OOD FAR-M case, FARM reaches the largest gain of $+46.2\%$, exceeding both Transformer ($+42.2\%$) and CRA Transfer ($+35.6\%$). 
These results indicate that the Stage-I reward prior captures trajectory-level evaluation structure that remains useful beyond the exact source-task configurations. 
As the distribution shift increases, the transfer benefit decreases but remains positive. 
On the hard open-set users $u_{15}$ and $u_{23}$, FARM achieves a mean improvement of $+21.6\%$, compared with $+18.1\%$ for the Transformer variant and $+6.4\%$ for CRA Transfer. 
Together with the target-task adaptation trajectories in Fig.~\ref{fig:stage2-transfer}(a), this trend indicates that transfer effectiveness depends on the compatibility between the learned reward knowledge and target-task dynamics. 
Tasks closer to the learned trajectory and objective structure obtain larger gains, whereas stronger distribution shifts reduce reusable reward information.

Overall, Stage-II results show that FARM improves target-policy adaptation
across seen and unseen task settings by reusing a frozen trajectory-level
reward prior without transferring source policies. Both Mamba and Transformer
support Reward-Space Transfer at the matched $W=32$ setting, indicating that
the mechanism is not tied to a specific sequence backbone. Longer-history
behavior is examined in \textbf{RQ3}.

\begin{table}[!t]
\centering
\small
\setlength{\tabcolsep}{3.5pt}
\caption{Stage-II late-stage gain over Single-task SAC (\%).
Boldface denotes the best result in each transfer setting.}
\label{tab:s2-agg}
\begin{tabular}{@{}lccc@{}}
\toprule
Scenario & CRA & FARM & TF \\
\midrule

Seen ($\tau_2,\tau_9$)
& $+1.2$ & $+10.0$ & $\mathbf{+10.2}$ \\

Unseen Rate$\wedge$Latency
(FAR-M, $u_{15}$, $u_{23}$)
& $+16.1$ & $\mathbf{+29.8}$ & $+26.1$ \\

Moderate-OOD (FAR-M)
& $+35.6$ & $\mathbf{+46.2}$ & $+42.2$ \\

Hard open-set ($u_{15}$, $u_{23}$)
& $+6.4$ & $\mathbf{+21.6}$ & $+18.1$ \\

\bottomrule
\end{tabular}
\end{table}












\begin{figure}[!t]
\centering
\includegraphics[width=0.95\linewidth]
{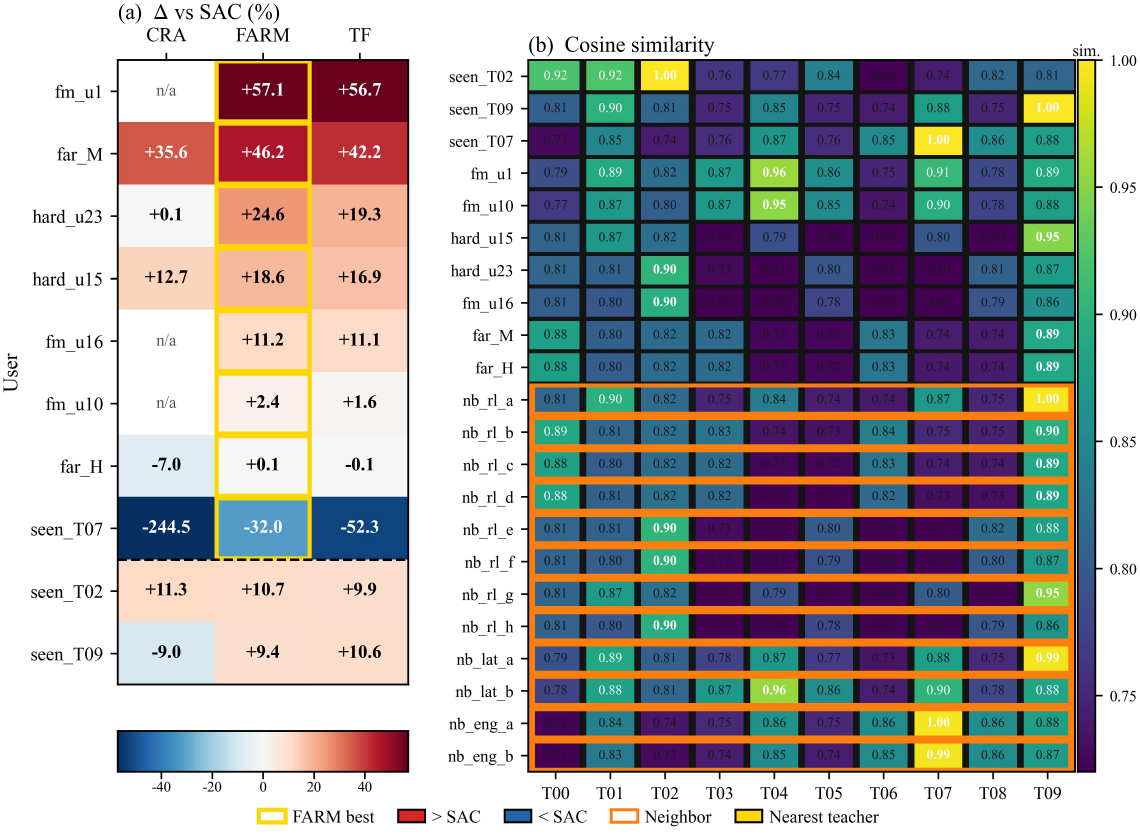}
\caption{
Per-user Stage-II transfer and task-condition similarity.
\textbf{(a)} Late-stage gain over scratch SAC for CRA, FARM
(Mamba-$L{=}2$), and Transformer-$L{=}2$; gold boxes mark the largest gain,
while red/blue cells indicate positive/negative transfer.
\textbf{(b)} Cosine similarity between Stage-II task conditions and the
Stage-I task bank; bold entries identify the nearest source task and orange
boxes denote constructed corridor neighbors. 
}
\label{fig:farm-ac}
\end{figure}

\begin{figure}[!t]
\centering
\includegraphics[width=0.95\linewidth]
{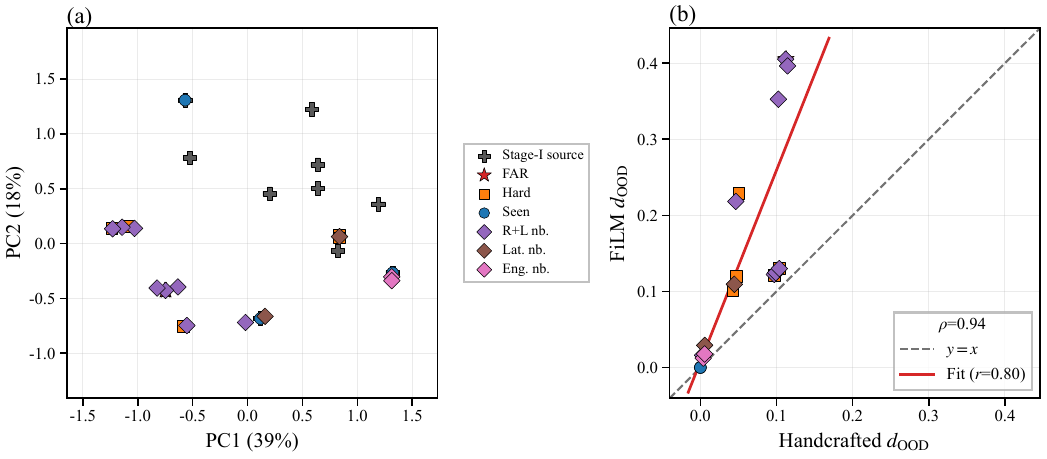}
\caption{
Stage-II task-condition geometry and OOD-distance consistency.
\textbf{Left:} PCA projection of task-condition vectors, with gray crosses
denoting Stage-I source tasks and colored markers denoting Stage-II targets.
\textbf{Right:} Handcrafted versus Constraint-FiLM OOD distances; the solid line the least-squares fit
(Pearson $r=0.80$, Spearman $\rho=0.94$).
}
\label{fig:farm-c2}
\end{figure}

\subsubsection{Main Transfer Gains}

Stage-II transfer is evaluated using the late-stage gain
$\Delta_{\mathrm{SAC}}$ defined in Section~\ref{sec:exp-report}. CRA Transfer,
FARM with the frozen Mamba-$L{=}2$ reward prior, and the matched
Transformer-$L{=}2$ variant use the same target-task interaction budget and
evaluation configuration.

Fig.~\ref{fig:stage2-transfer}(b) and Fig.~\ref{fig:stage2-family} summarizes the transfer results from two complementary
perspectives. Fig.~\ref{fig:stage2-transfer}(b) relates transfer gain to target-task
distribution shift, characterized by the distance to the nearest Stage-I task
and the task-difficulty score. Across the measured open-set users and
constructed corridor neighbors, FARM yields positive transfer in most
evaluated cases and generally larger gains than CRA Transfer, while the
Transformer variant remains competitive. Fig.~\ref{fig:stage2-family} aggregates the results by user family. For the
unseen Rate--Latency group consisting of FAR-M, $u_{15}$, and $u_{23}$,
FARM achieves a mean gain of $+29.8\%$, compared with $+26.1\%$ for
Transformer-$L{=}2$ and $+16.1\%$ for CRA Transfer. On the in-distribution
controls T02 and T09, FARM and Transformer achieve similar gains. 

\subsubsection{Per-User Transfer Analysis}

Fig.~\ref{fig:farm-ac}(a) further reports the Stage-II results at the
user level. Among users with complete three-way comparisons, FARM achieves
the largest gain in most cases, including the moderate-OOD case FAR-M
($+46.2\%$) and the hard Rate--Latency users $u_{15}$ and $u_{23}$.
These user-level results are consistent with the aggregate results in
Fig.~\ref{fig:stage2-transfer}(b), indicating that the overall improvement is
not driven by a small subset of favorable targets. For the seen cases, FARM
and Transformer remain close, suggesting that backbone differences are less
pronounced when the target configuration is already represented in the
Stage-I task bank. The seen energy case corresponding to T07 is retained in
the heatmap for completeness but excluded from the task-geometry scatter
because its percentage gain over Single-task SAC exceeds the plotting range.

\subsubsection{Stage-II Evaluation-Set Geometry}

To relate transfer behavior to the structure of the target-task distribution,
Fig.~\ref{fig:farm-ac}(b) and Fig.~\ref{fig:farm-c2} characterize the geometry
between Stage-II users and the Stage-I task bank. Fig.~\ref{fig:farm-ac}(b) reports the cosine similarity between each Stage-II
task and the source-task bank. Seen users coincide with their corresponding
source tasks, whereas FAR and hard open-set users remain closest to related
source configurations but exhibit lower similarity. The evaluation set
therefore represents controlled degrees of distribution shift rather than
unstructured task perturbations. Fig.~\ref{fig:farm-c2} further examines whether this externally defined task
geometry is consistent with FARM's internal condition representation. The left
panel shows that held-out users occupy structured neighborhoods around the
Stage-I source tasks in the original condition space. The right panel compares
$d_{\mathrm{OOD}}^{\mathrm{hand}}$ with the corresponding
$d_{\mathrm{OOD}}^{\mathrm{FiLM}}$ after the Constraint-FiLM mapping.
Their strong agreement (Pearson $r=0.80$, Spearman $\rho=0.94$) indicates
that the relative near/far relationships used to characterize the evaluation
set are largely preserved by FARM's condition pathway. Thus, Fig.~\ref{fig:stage2-transfer}(b) and Fig.~\ref{fig:stage2-family} and~\ref{fig:farm-ac}(a) provide the primary
performance evidence for Stage-II transfer, while
Figs.~\ref{fig:farm-ac}(b) and~\ref{fig:farm-c2} support the construction and
interpretation of the corresponding distribution shifts.

\subsection{Temporal Representation for Reward Prior Transfer}
\label{sec:exp-rq3}

\begin{figure}[!t]
\centering
\includegraphics[width=0.6\linewidth]
{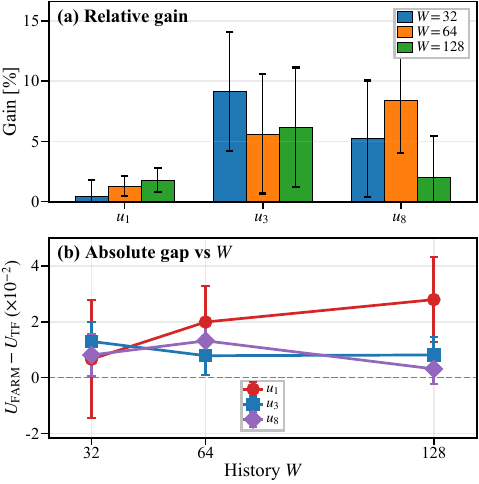}
\caption{
Temporal-encoder comparison on users with unseen tasks $u_1$, $u_3$, and $u_8$.
Mamba-$L{=}2$ and Transformer-$L{=}2$ reward priors are pretrained with
$W=32$ and frozen during Stage-II adaptation.
\textbf{Left:} relative late-stage gain of Mamba over Transformer,
$(U_{\mathrm{FARM}}-U_{\mathrm{TF}})/|U_{\mathrm{TF}}|$, for
$W\in\{32,64,128\}$.
\textbf{Right:} absolute late-stage objective gap
$U_{\mathrm{FARM}}-U_{\mathrm{TF}}$ versus history length.
Error bars are obtained from the late-window standard deviations using
$\sigma_\Delta=
\sqrt{\sigma_{\mathrm{FARM}}^2+\sigma_{\mathrm{TF}}^2}$,
with the left panel additionally normalized by $|U_{\mathrm{TF}}|$.
The absolute gaps are shown on a $\times10^{-2}$ scale.
Positive values indicate better performance with Mamba.
}
\label{fig:w-sweep-mamba}
\end{figure}

We finally investigate \textbf{RQ3}: how the temporal trajectory encoder
affects Reward-Space Transfer when the inference history extends beyond the
Stage-I training configuration. We compare the default Mamba-$L{=}2$ encoder
with a matched causal Transformer-$L{=}2$ encoder. Both reward priors use the
same hidden dimension, objective-grouped experts, FiLM conditioning, and
CondLam modules and are pretrained with $W=32$. After Stage-I training, their
parameters are frozen and transferred through ObjRoute AFT, while the
condition-dependent modules continue to be evaluated using the corresponding
target-task condition during adaptation. The evaluation uses three randomly generated held-out users,
$u_1$, $u_3$, and $u_8$, drawn from the Stage-II open-set generator. Their
channel characteristics, temporal correlations, traffic conditions, and QoS
configurations form combinations not jointly observed in the Stage-I bank.
The inference history is varied over $W\in\{32,64,128\}$: $W=32$ matches the
Stage-I training condition, whereas $W=64$ and $W=128$ test whether the frozen
reward prior can exploit longer trajectory histories without additional
pretraining. To reduce the influence of early exploration, performance is
measured using the mean $U_{\tau}$ over the final $15\%$ of the adaptation
trajectory. For each user and history length, let $U_{\mathrm{FARM}}$ and
$U_{\mathrm{TF}}$ denote the corresponding late-window mean returns, and let
$\sigma_{\mathrm{FARM}}$ and $\sigma_{\mathrm{TF}}$ denote their late-window
standard deviations. Assuming the two late-window fluctuations are independent,
the uncertainty of the absolute gap is propagated as
$
\sigma_{\Delta}
=
\sqrt{
\sigma_{\mathrm{FARM}}^{2}
+
\sigma_{\mathrm{TF}}^{2}
}.
$
Accordingly, the error bar for the relative gain is
$100\sigma_{\Delta}/|U_{\mathrm{TF}}|$.
Fig.~\ref{fig:w-sweep-mamba} reports both the relative gain of Mamba over
Transformer and their absolute objective gap.

Under the original history length $W=32$, Mamba and Transformer provide
comparable transfer performance, with Mamba retaining a positive relative
gain for all three users. The larger gains occur for $u_3$ and $u_8$, at
approximately $9\%$ and $5\%$, respectively. This result shows that both
temporal encoders can support effective Reward-Space Transfer when the
inference history matches the Stage-I training condition.

The difference becomes more apparent as the history length increases.
For $u_1$, the absolute late-stage gap increases from approximately $0.007$
at $W=32$ to $0.020$ at $W=64$ and $0.028$ at $W=128$.
Mamba also maintains a positive gap over Transformer for $u_3$ and $u_8$
across all tested history lengths. Since the corresponding returns lie near
$U\approx-1.5$ to $-0.13$, the absolute gaps are on the order of
$10^{-2}$, consistent with the $\times10^{-2}$ scale used in the right panel. These results suggest that selective state-space modeling provides additional
robustness when the frozen reward prior must process longer temporal contexts
than those observed during Stage-I training. Overall, both temporal backbones
support the FARM transfer mechanism under the matched $W=32$ configuration,
while the Mamba-based ARM becomes more robust as the available trajectory
history increases. The observed advantage is therefore associated primarily
with extended temporal representation rather than a universal improvement
across all operating conditions. Based on this comparison, FARM uses
Mamba-$L{=}2$ as the default trajectory encoder and retains
Transformer-$L{=}2$ as the matched architectural ablation.

\section{Conclusion}
\label{conclusion}

This paper presented FARM, a transferable reward-prior framework for
heterogeneous wireless optimization. FARM learns trajectory-level evaluation
knowledge through ARM and reuses the learned reward prior to guide
task-specific policy optimization, enabling Reward-Space Transfer without
transferring source-task policies. Experiments show that FARM reduces the interference associated with direct
policy sharing in Stage~I and improves Stage~II adaptation to seen and unseen
task configurations over independent SAC and conventional shared-reward
transfer. Both Mamba and Transformer support reward-prior transfer, while
Mamba exhibits robustness under longer trajectory histories. These
results demonstrate the effectiveness of reward-space knowledge reuse for
heterogeneous wireless control. Future work will consider larger-scale,
continually evolving, and cross-domain wireless environments.

\bibliographystyle{IEEEtran}
\bibliography{Ref}

\end{document}